# Resistive Switching Characteristics of Al/Si$_3$N$_4$/*p*-Si MIS-Based Resistive Switching Memory Devices

**Min Ju Yun, Sungho Kim\*, and Hee-Dong Kim\*\***

*Department of Electrical Engineering, Sejong University, Neungdong-ro 209, Gwangjin-gu, Seoul 143-747, Korea*

In this study, we proposed and demonstrated a self-rectifying property of silicon nitride (Si$_3$N$_4$)-based resistive random access memory (RRAM) device by employing *p*-type silicon (*p*-Si) as bottom electrode. The RRAM devices consisted of Al/Si$_3$N$_4$/*p*-Si are fabricated by a low-pressure chemical vapor deposition and exhibited an intrinsic diode property with non-linear current−voltage (*I*−*V*) behavior. In addition, compared to conventional metal/insulator/metal (MIM) structure of Al/Si$_3$N$_4$/Ti RRAM cells, operating current in whole bias regions for proposed metal/insulator/semiconductor (MIS) cells has been dramatically lowered because introduced *p*-Si bottom electrode efficiently suppresses the current in both low and high resistive states. As a result, the results mean that by employing *p*-Si as bottom electrode the Si$_3$N$_4$-based RRAM cells can be applied to selector-free RRAM cells.

PACS number: 73.40.Qv, 68.55.Jk

Keywords: RRAM; Si$_3$N$_4$ Films; Self-rectifying

\*E-mail: sungho85.kim@sejong.ac.kr, tel: +82-2-6935-2422, fax: +82-2-3408-4329

\*\*E-mail: khd0708@sejong.ac.kr, tel: +82-2-6935-2423, fax: +82-2-3408-4329

# I. INTRODUCTION

Currently, many researchers have been tried to explore new memory concept, i.e., two terminal resistance change-based memories since the size of conventional capacitance-based nonvolatile memory (NVM) memory cells will be approaching their scaling limits [1,2]. Among various kinds of new memory concepts, resistive switching random access memory (RRAM) consisted of various metal oxides and metal nitrides have been proposed as one of the most promising candidates because of its simple metal/insulator/metal (MIM) structure, low-voltage and high-speed operation, and high density features [3,4]. Nevertheless, some issues in respect of reliability should be solved to be realized as new NVMs in terms of unit-cell and array-cell levels. First, in unit cell, a current level for SET and RESET operations should be lowered below sub-micro ampere to reduce a power consumption and improve a device's reliability such as endurance by optimizing an active layer and structures of RRAM devices [5]. That is, so as to reduce a variation of operating parameters, more reliable deposition method to get uniform resistive switching (RS) films, like atomic layer deposition (ALD) or chemical vapor deposition (CVD), should be employed on behalf of physical vapor deposition (PVD) since achieving uniform RS films is very difficult on whole wafer using conventional PVD method [6]. Moreover, in array, the disturbance (or crosstalk) induced at junctions between the neighboring cells should be controlled since the leakage path formed due to reversely biased cells [7]. In relevant work experience some kinds of selectors, for example, extrinsic diodes in unipolar mode cells or transistors combined with bipolar RS (BRS) cells, have been investigated to solve the issue [8,9], which is so called one diode–one resistor (1D1R) or one transistor–one resistor (1T1R). However, the 1D1R and 1T1R structures with the increased feature cell sizes in an array might be a shortcoming in terms of device's integration as well as mass production. Therefore, in order to protect the stored data by leakage current during reading operation and increase the current read-margin, RRAM architectures having self-selector function should be realized.

To capture BRS cells into array structure, in this work, we investigate RS behavior with current limiter of an Al/Si$_3$N$_4$/*p*-Si structures for the applications of RRAM using CVD method that can provide a reliable film quality of active layer, compared to conventional PVD method. In this structure, a Si$_3$N$_4$ layer carries out BRS behaviors while *p*-Si bottom electrode acts as bottom electrode as well as the selector (or current limiter) in the positive bias region. The electrical properties as well as read margin values were investigated.

## II. EXPERIMENTS AND DISCUSSION

For fabricating the metal-insulator-silicon (MIS) sample preparation, (1 0 0)-oriented *p*-type silicon (*p*-Si) wafers were used as the starting substrates, which acts as bottom-electrode in this system. A 8-nm-thick Si$_3$N$_4$ layer was subsequently deposited by low pressure CVD (LPCVD) at 750 °C by the reaction of dichlorosilane (SiCl$_2$H$_2$) and ammonia (NH$_3$) gas. And then, cleaning process of the Si$_3$N$_4$/*p*-Si structure was performed using a standard sulfuric acid and hydro-peroxide mixture (SPM) for 10 min. The thicknesses of the Si$_3$N$_4$ stacks were confirmed to be 8-nm by spectroscopic ellipsometry and α-step as well. Then, the X-ray diffraction (XRD) for the Si$_3$N$_4$ films to reveal its structural properties was measured. Figure 1(c) exhibits that the Si$_3$N$_4$ films have a face-centered cubic crystal Structure, where a marked diffraction peak observed at 36.5°, 42.2° and 53.6° originates 220-, 311- and 400-planes, in the scanned range of 30°-60°. To confirm the roughness of Si$_3$N$_4$ films, the surface morphology of the Si$_3$N$_4$ films was also observed using atomic force microscopy (AFM), as shown in Fig. 1(d). The AFM micrograph shows that the Si$_3$N$_4$ films were uniformly deposited with an root-mean-square (RMS) surface roughness of ~4 nm over a 500 x 500 nm$^2$ area. After drying in a nitrogen gas, we continually deposited a 100 nm thick top Al electrode with a 100 μm diameter by using a sputtering system and a bottom indium (In) contact was made on the substrate to make an Ohmic contact. Figure 1(a) shows the finished structures of proposed Al/Si$_3$N$_4$/*p*-Si cells. In addition, in order to compare the MIS based RRAMs with the MIM based RRAMs, we additionally fabricated

Al/Si$_3$N$_4$/Ti structures, as shown in Fig. 1(b). The electrical properties of the RRAM samples were measured using a Keithley 4200 semiconductor parameter analyzer.

First, to study the BRS properties of the Al/Si$_3$N$_4$/*p*-Si (MIS) and Al/Si$_3$N$_4$/Ti (MIM) RRAM cells, the dc current−voltage (*I−V*) characteristics of them were investigated at room temperature. In the Al/Si$_3$N$_4$/Ti RRAM cells, at virgin state (i.e. HRS) we performed the forming process in positive bias region and it was observed at +7 V. Then, the BRS is achieved by sweeping dc *I−V* in the following sequence: 0 V → −1 V → 0 V → +1 V → 0 V, as shown in Fig. 2(a). A SET process from the HRS to the LRS is performed at around +0.7 V with a compliance current (CC) of 100 mA. For example, the RRAM cell can be switched from the HRS to the LRS. Subsequently, the current of the cell gradually drops at about around −0.7 V by sweeping the bias voltage in the negative voltage region, indicating that the state is switched back from the LRS to HRS. On the other hand, the MIS samples also play BRS behavior in the dc *I−V* curve in the following sequence: 0 V → −15 V → 0 V → +15 V → 0 V, as shown in Fig. 2(b). As a result, when compared with the MIM sample, in the negative bias region, current at the LRS lowered from ~30 mA (or 3.82 MA/m$^2$) at 0.2 V to13 µA (or 1.65 kA/m$^2$) at 1 V, and the current at the HRS decreased from ~8 mA (or 1.01 MA/m$^2$) to ~30 pA (or 3.82 mA/m$^2$), at 1 V, respectively. Therefore, compared to the current ratio (CR) between the HRS and the LRS of two samples, the increased CR of <10$^6$ was observed in the MIS structure, as shown in Figs. 1(a) and (b). That might be due to current suppression effect by using *p*-Si bottom electrode. In addition, the current is limited under whole positive bias region and we have also achieved non-linear *I−V* curve, leading to asymmetric *I–V* curve. Especially, during the voltage sweep in positive bias to RESET the device, any RS behavior could not be happened, which shows that the proposed MIS structure plays an intrinsic rectifying property. It can suppress the positive bias current like the diode even if the conduction paths in the LRS is formed. In addition, compared to current level of two samples, we have observed similar current levels at the LRS while in this test the big difference was formed at the HRS, which might be explained that relatively more conducting filaments (CFs) in the active layer of the MIM samples than that of the MIS samples were generated via the SET process. As a result, even after RESET process it might be possible to exist some CFs in the active layer and higher carrier transfer through them can be induced at the HRS, compared to the MIS samples.

Then, so as to study a conducting mechanism for the MIS $Si_3N_4$-based memory cell, we have tried to replot the dc $I-V$ curves using some models, such as space charge limited current (SCLC), Ohmic behavior, and Poole-Frenkel current ($I_{P-F}$) equation, according to the related literatures [8-10]. From the result, it was found that the curves at the HRS and LRS were well matched with SCLC, as shown in Fig. 3. In this figure, we used a log-log scale in order to reveal the power law relation ($I \propto V^m$), as can be seen in Fig. 3. In the negative voltage region, the slopes of the HRS and the LRS are about 1 in the low voltage region, which corresponds to the Ohm's law ($I \propto \sim V^1$). And then, the currents of both the HRS and LRS follow the square dependence on the voltage, corresponding to the Child's square law ($I \propto \sim V^2$). Finally, the current of the HRS increases very quickly at trap filled limit voltage ($V_{TFL}$), corresponding to a steep increase with the increase of the voltage. The carrier transport behavior is well consistent with the modified SCLC mechanism incorporating the Poole-Frenkel effect [8]-[10]. Consequently, the result shows that the basic switching mechanism of $Si_3N_4$-based RRAM is closely related to the electron trapping and de-trapping processes in nitride-related electron traps or dangling bonds. For example, the current path can be formed within the $Si_3N_4$ films through the trap-to-trap hopping process of the electrons (SET process) if the majority of the electron traps are occupied. By contrast, if the majority of the electron traps are empty, the current path is broken down via the electron de-trapping process (RESET process).

Finally, in order to suppress a read interference by the parasitic sneak current path in the CBA structures, we employed the MIS-based RRAM devices, consisted of the Al/$Si_3N_4$/$p$-Si structures. In the $I$-$V$ curve characteristics, we observed its nonlinear selection behavior (at low voltage region in both biases) as well as Schottky diode property with current rectifying in positive bias region, as shown in Fig. 2(b). Therefore, the intrinsic current limit behavior in the positive bias regions can suppress disturbance between adjacent elements under a positive bias during reading process. To further discuss the selector property of the proposed RRAM devices, we evaluated on the read margin voltage ($\Delta V$) normalized to the pull-up voltage ($V_{pu}$) by making out the Kirchhoff equation, as shown in Fig. 4. As a result, the MIM $Si_3N_4$ RRAM makes an interference phenomena by sneaky leakage path over 2 × 2 CBA, while the proposed MIS structure covers 17x17 arrays during read operating due to its current limit property in low voltage.

## III. CONCLUSION

In summary, to integrate bipolar resistive switching (BRS) cells into array structures, we proposed Al/$Si_3N_4$/*p*-Si MIS structures for the applications of array RRAM and demonstrated a feasibility of self-rectifying resistive switching behavior in this cell. As a result, we observed an asymmetric $I-V$ curve in positive and negative bias regions as well as non-linear property in $I-V$ curve, while using MIS structures, the read margins abruptly increased from 4 to 300, compared to MIM structures. Therefore, this device can potentially simplify the fabrication process in high-density array applications.

## ACKNOWLEDGEMENT

H.-D. Kim and M.J. Yun contributed equally to this work. This research was supported by Basic Science Research Program through the National Research Foundation of Korea (NRF) funded by the Ministry of Education (No. 2015R1D1A1A01056803).

## REFERENCES

[1] A.B.K. Chen, B.J. Choi, X. Yang, and I.W. Chen, Adv. Funct. Mater. **22**, 546 (2011).

[2] L. Chen, Y. Xu, Q.Q. Sun, H. Liu, J.J. Gu, S.J. Ding, and D.W. Zhang, IEEE Electron Device

Lett. **31**, 356 (2010).

[3] Y.C. Bae, A.R. Lee, J.B. Lee, J.H. Koo, K.C. Kwon, J.G. Park, H.S. Im, and J.P. Hong, Adv. Funct. Mater. **22**, 709 (2011).

[4] H.D. Kim, M.J. Yun, and T.G. Kim, Phys. Status Solidi-Rapid Res. Lett. **6**, 81 (2015).

[5] C. Walczyk, D. Walczyk, T. Schroeder, T. Bertaud, M. Sowinska, M. Lukosius, M. Fraschke, D. Wolansky, B. Tillack, E. Miranda, and C. Wenger, IEEE Transaction on Electron Devices, **58**, 3124 (2011).

[6] S.S. Lee, E.S. Lee, S.H. Kim, B.K. Lee, S.J. Jeong, J.H. Hwang, C.G. Kim, T.M. Chung, and K.S. An, Bull. Korean Chem. Soc. **33**, 2207 (2012).

[7] E. Linn, R. Rosezin, C. Kügeler, and R. Waser, Nat. Mater. **9**, 403 (2010).

[8] T. Bertaud, D. Walczyk, M. Sowinsk, D. Wolansky, B. Tillack, G. Schoof, V. Stikanov, C. Wenger, S. Thiess, T. Schroeder and C. Walczyk, ECS Trans. **50**, 21 (2013).

[9] H.D. Kim, M.J. Yun, and T.G. Kim, Appl. Phys. Lett. **105**, 213510 (2015).

[10] L. Tang, P. Zhou, Y.R. Chen, L.Y. Chen, H.B. Lv, T.A. Tang and Y.Y. Lin, J. Korean Phys. Soc. **53**, 2283 (2008).

Figure Captions.

Fig. 1. Fabricated RRAM structures of (a) Al/Si$_3$N$_4$/Ti MIM and (b) proposed self-rectifying RRAM cells with Al/Si$_3$N$_4$/*p*-Si structures. (c) The XRD curves measured for the Si$_3$N$_4$ films in the scanned range of 30°−60°. (d) Typical AFM topography over a 1 x 1 µm$^2$ area for the Si$_3$N$_4$ films.

Fig. 2. *I–V* characteristics of (a) the Al/Si$_3$N$_4$/Ti structure and (b) the proposed Al/Si$_3$N$_4$/*p*-Si structure in linear scale.

Fig. 3. Space charge limited conduction (SCLC) model with ln [bias voltage] vs. ln [current] for HRS and LRS.

Fig. 4. Read margin ΔV/V$_{pu}$ as a function of number of word lines in crossbar configurations.

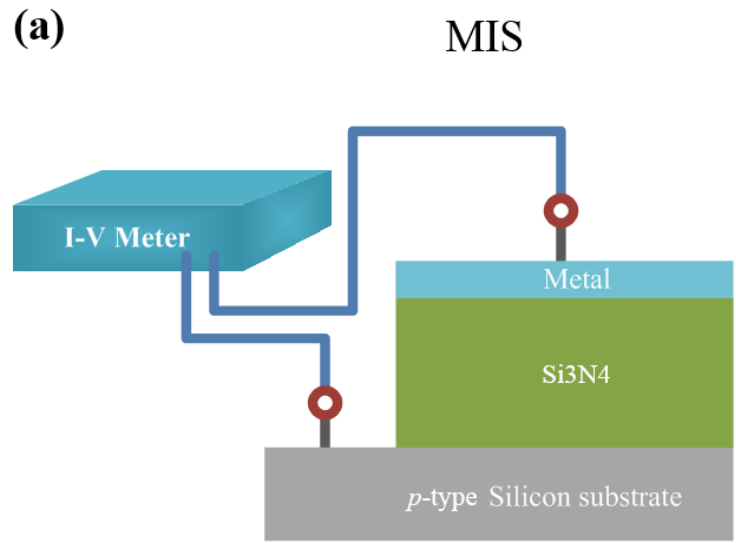

Figure 1(a)

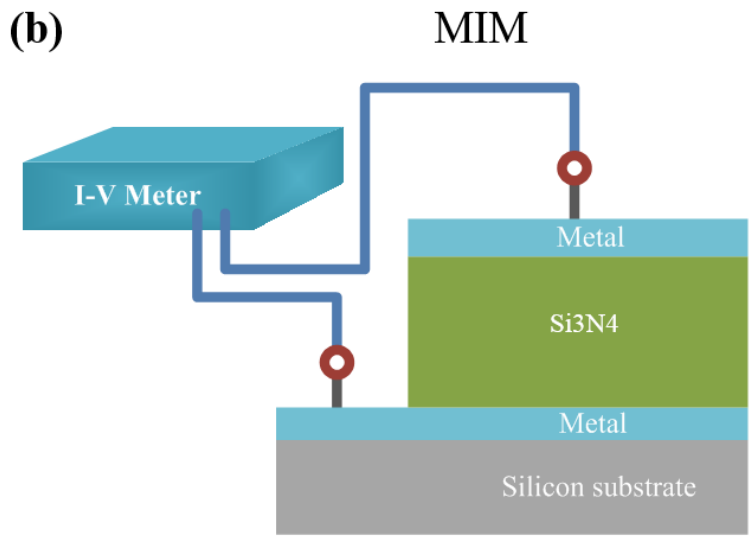

Figure 1(b)

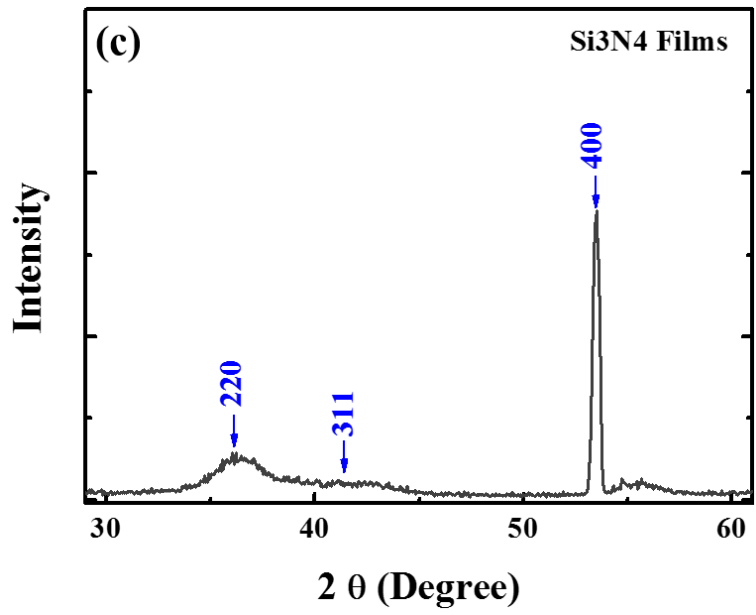

Figure 1(c)

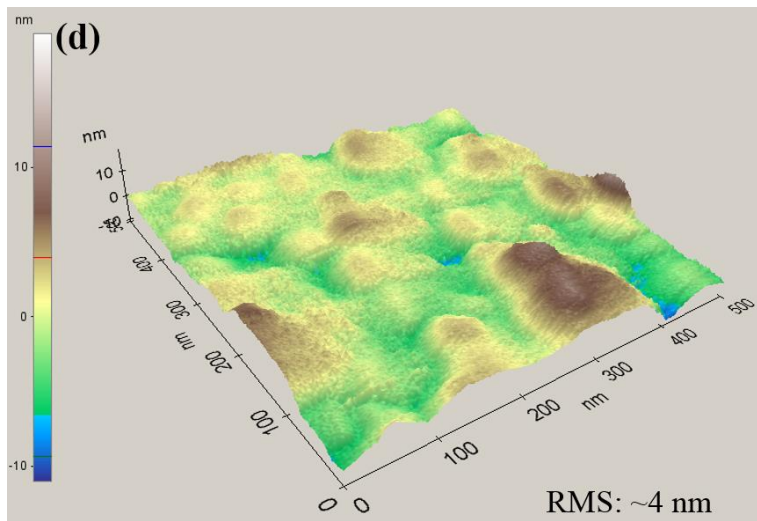

Figure 1(d)

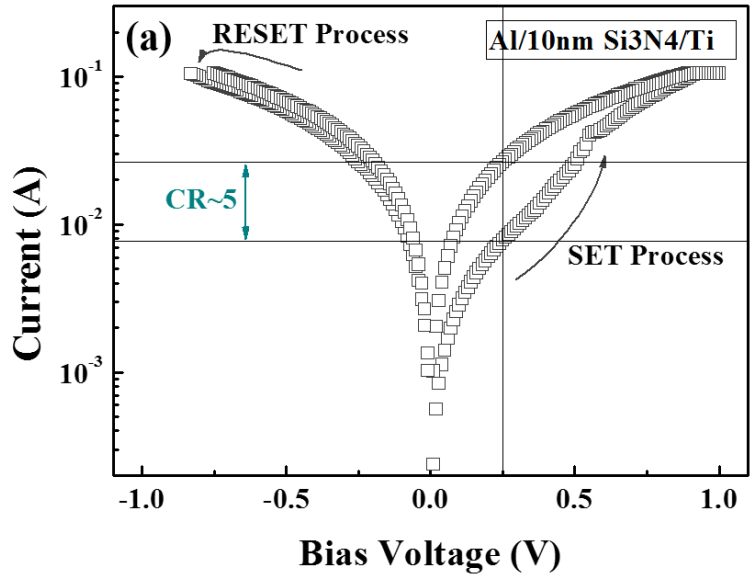

Figure 2(a)

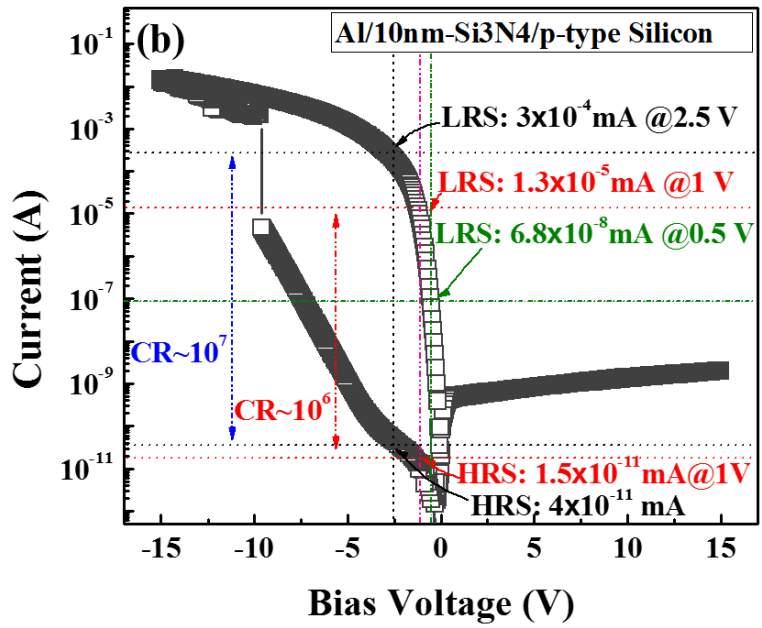

Figure 2(b)

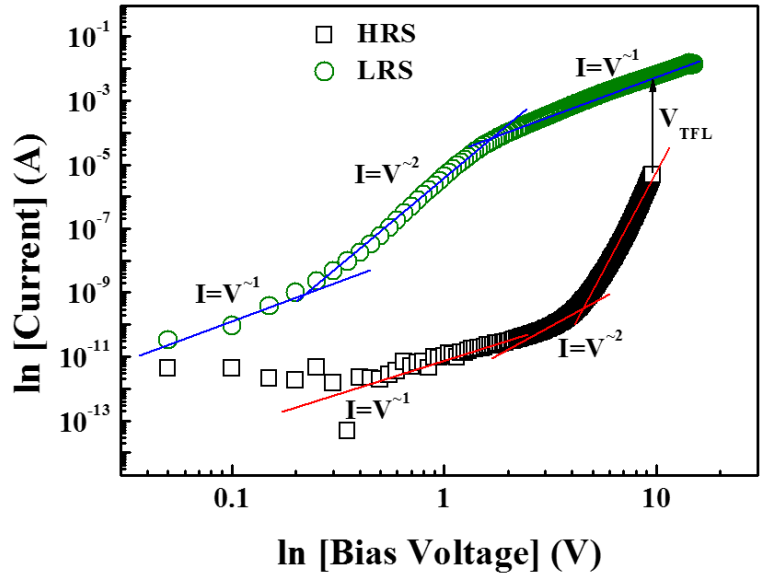

Figure 3

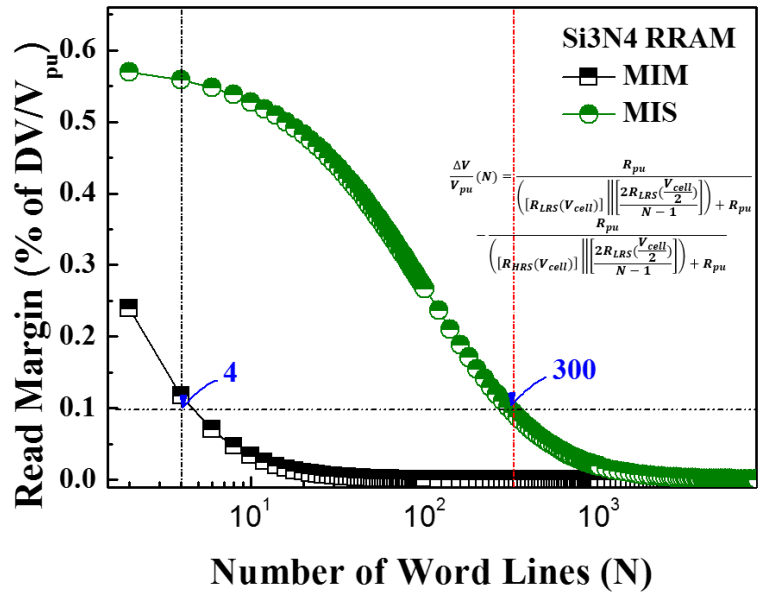

Figure 4